\documentclass[prl, twocolumn, showpacs]{revtex4}

\usepackage[utf8]{inputenc}
\usepackage[english]{babel}
\usepackage{graphicx}
\usepackage{amssymb}

\begin{document}

\title{Scale-free phase field theory of dislocations}

\author{Istv\'an Groma}
\email{groma@metal.elte.hu}
\affiliation{Department of Materials Physics, E\"otv\"os University Budapest,
H-1517 Budapest POB 32, Hungary}

\author{Zolt\'an Vandrus}
\affiliation{Department of Materials Physics, E\"otv\"os University Budapest,
H-1517 Budapest POB 32, Hungary}

\author{P\'eter Dus\'an Isp\'anovity}
\affiliation{Department of Materials Physics, E\"otv\"os University Budapest,
H-1517 Budapest POB 32, Hungary}

\begin{abstract} 
According to recent experimental and numerical investigations if the characteristic size of a specimen is 
in the submicron size regime several new interesting phenomena emerge during the deformation of the samples. 
Since in such a systems the boundaries play a crucial role, to model the plastic response of 
submicron sized crystals it is crucial to determine the dislocation distribution near the boundaries. 
In this paper a  phase field type of continuum theory of the time evolution of an ensemble of 
parallel edge dislocations with identical Burgers vectors, corresponding to the 
dislocation geometry near boundaries, is presented. Since the 
dislocation-dislocation interaction is scale free ($1/r$), apart from the average 
dislocation spacing the theory cannot contain any length scale parameter. As 
shown, the continuum theory suggested is able to recover the dislocation distribution near 
boundaries obtained by discrete dislocation dynamics simulations. 

\end{abstract}

\pacs{62.25.-g, 61.72.Lk, 64.70.qj, 45.70.Ht}
\maketitle

Three dimensional crystals \cite{nabarro1967theory} and different 2D lattices, 
like Abrikosov vortices 
\cite{hellerqvist1996vortex,moretti2009irreversible}, 
charge density waves \cite{gill1996dislocations,feinberg1988elastic},  or Wigner 
solids \cite{monarkha2004two}, generically contain a large number of line-type topological  defects called 
dislocations greatly affecting the plastics response of these  systems.  Thus 
studying the collective  properties of interacting dislocations is of utmost 
relevance in material physics. Although the interaction and dynamical properties 
of individual dislocations are well known for a long time,    in most cases the 
deformation properties of the crystalline materials are controlled by the 
collective evolution of a large number of dislocations. One approach to model 
the rather complex phenomena caused by the collective motion of dislocations is 
the numerical solution of the equation of motion of individual dislocations 
called discrete dislocation dynamics (DDD). During the past decades numerous DDD 
simulation algorithms have been developed both in 2 
\cite{nicola2005size,needleman2003discrete,
miguel2001, miguel2002,gomez2006dislocation,needleman2006,bako2007dislocation,benzega2008,
guruprasad2008phenomenological,ispanovity2010submicron,ispanovity2011criticality}
and 3 \cite{madec2002,wang2006,devince2008,weygang2008,senger2011} dimensions, allowing to study problems 
like hardening \cite{madec2002,benzega2008,devince2008}, size effect 
\cite{balint2006,guruprasad2008phenomenological,weygang2008,benzerga2009,senger2011}, 
jamming-flowing transition \cite{miguel2002,ispanovity2014}, relaxation \cite{ispanovity2011criticality} 
dislocation avalanches \cite{miguel2001,csikor2007,ispanovity2010submicron}, etc. 

One may expect, however, that for a large number of problems not all the details 
accounted  by DDD simulations are important, the response of the dislocation 
network can be well described on a continuum level. Although several such 
continuum theories of dislocations have been developed,  
\cite{azab2000,levkovitch2006,acharya2006,sedlacek2007,hochrainer2007,kratochvil2008,limkumnerd2008,
roy2008,mesarovic2010,sandfeld2011,poh2013} 
most of them correspond either to mean field approximation or are based on completely phenomenological 
grounds. However, the role of dislocation-dislocation correlation, crucial 
because of the long range nature of dislocation-dislocation interaction, is far 
from understood. Correlation effects are taken into account in a systematic 
manner only in the limit when the signed dislocation density $\kappa$ (geometrically necessary 
dislocation (GND) density) is much smaller than the 
stored density $\rho$.
\cite{groma1997link,groma2003spatial,groma2006debye,groma2007dynamics,groma2010variational}.    

With the advance of nanotechnology the characteristic size of the microstructure 
of crystalline materials reduced to the submicron level. As a consequence, the 
role of boundaries (sample surface, grain boundary, etc.) has become even more 
important than earlier. So, to model the plastic response of samples with 
features on the submicron scale it is crucial to determine the dislocation 
distribution near the boundaries. Close to a boundary the 
GND density is often comparable to the stored one, so the assumption $|\kappa|\ll\rho$ 
is not valid. 

The dislocation distribution near a boundary is traditionally described by the 
1D pile-up of the dislocations \cite{hirsch}. For many real dislocation configurations, 
however, the interaction between dislocations in different slip planes is 
important requiring to go up to modeling in minimum of 2D. In this paper a phase 
field type  theory is suggested for the simplest possible 2D dislocation 
arrangement consisting of straight parallel dislocations with single slip. The 
evolution equations of the dislocation densities are obtained from a functional 
of the dislocation densities and the stress potential. In contrast to other 
approaches suggested recently, where a set of walls of dislocations with 
equidistant slip distances is considered to model the dislocation configuration 
near the boundary \cite{roy2008,poh2013}, here we assume that the slip planes of the dislocations 
are arranged completely randomly. (In the present model dislocation climb is 
excluded, so the dislocations cannot leave their slip planes). Because of the 
$1/r$, {\it i.e.} 
scale-free, nature of 
dislocation-dislocation interaction, a key consequence of the random slip plane 
setup is that beside the coarse grained local dislocation spacing no other 
parameter with a length scale  can appear in the theory.  As it is explained in 
detail below this scale-free nature largely determines the possible form of the 
phase field potential. We speculate that  the framework suggested could be 
applicable to other systems with scale free interaction, like gravitation.

Let us consider a system of parallel edge dislocations with line vectors 
$\vec{l}=(0,0,1)$ 
and Burgers vectors $\vec{b}_{\pm}=\pm(b, 0,0)$. The force in the slip plane 
acting on a dislocation is $b \tau$ where $\tau$ is the shear stress 
generated by the other dislocations plus the external shear.  It is commonly 
assumed that the velocity of a dislocation is proportional to the shear stress 
at the dislocation (over-damped dynamics) \cite{groma2003spatial}. So, the 
equation of the motion of the $i$th dislocation positioned at point $\vec{r}_i$ 
is  
\begin{equation}
\frac{dx_i}{dt}=Mb\tau(\vec{r}_i)=Mb_i\left(\sum_{j=1, j\ne i}^N s_j\tau_{\rm 
ind}(\vec{r}_i-\vec{r}_j)+\tau_{\rm ext} \right)
\label{eq_m}
\end{equation}
where  $M$ is the dislocation mobility, $\tau_{\rm ind}$ is the stress field 
generated by a dislocation, $\tau_{\rm ext}$ is the external stress, and 
$s_i=b_i/b=\pm 1$. The coupled system of equations of motion can be solved numerically,
that is called discrete dislocation dynamics (DDD) simulation. 

As it was shown in detail in \cite{groma2006debye,groma2007dynamics,groma2010variational} 
the equation of motion of the dislocations Eq. 
(\ref{eq_m}) can be obtained from the variational ``plastic'' potential 
\begin{equation}
 P^d[\chi,\rho^d]=\int \left[-\frac{D}{2}(\vartriangle \chi)^2+b\chi \partial_y 
(\rho^{d}_+-\rho^d_-)\right]{\rm d}x{\rm d}y
 \label{eq_Pd}
\end{equation}
as
\begin{equation}
 \frac{\delta P^{\rm d}}{\delta \chi}=-D\vartriangle^2 \chi+b\partial_y 
(\rho^{d}_+-\rho^d_-)= 0, \ \ \dot{\vec{r}}_i=M \vec{b}_i  \frac{\partial P^{\rm 
d}}{\partial \vec{r}_i} \label{eq_motion}
\end{equation}
where $D$ is a constant depending on the elastic moduli, $\chi$ is the stress 
function with $\tau=\partial_x\partial_y\chi$, and 
$\rho^d_{\pm}(\vec{r})=\sum_{i=1}^{N_{\pm}}\delta(\vec{r}-\vec{r}_i)$ in which 
the summation has to be taken for the positive or negative signed dislocations, 
respectively. So, $\rho^d_{+}(\vec{r})$ and $\rho^d_{-}(\vec{r})$ are the  
``discrete'' dislocation densities with the corresponding signs. 

One may expect, however, that for many problems not all the details represented 
by the discrete description are needed. So, with appropriate coarse-graining one 
can obtain a continuum theory suitable to model the evolution of inhomogeneous 
dislocation systems. In order to derive a continuum theory from the ``discrete'' 
evolution equation, as a first step, one can replace in $P^{\rm d}$ given by Eq. 
(\ref{eq_Pd}) the ``discrete'' $\rho^{\rm d}_{\pm}$ fields by their local 
averages $\rho_{\pm}$, leading to the form 
\begin{equation}
 P_{\rm sc}[\chi,\rho_{\pm}]=P^{\rm d}[\chi,\rho_{\pm}].
 \label{eq_Pm} 
\end{equation}
 Although by applying the standard formalism of phase field theories, from 
$P_{\rm sc}$ one can derive evolution equations for the fields $\rho_{\pm}$  in 
a systematic manner (see below), as it is explained in detail in \cite{groma2006debye,groma2007dynamics},
$P_{\rm sc}$ corresponds to  the mean (self-consistent) field approximation, i.e., 
dislocation-dislocation correlation effects are completely neglected. Due to the 
     long-range nature of dislocation-dislocation interaction correlation 
effects are extremely important. So, terms accounting for correlations have to 
be added to $P_{\rm sc}$ to arrive at a physically relevant model. 

As it is explained in detail in \cite{groma2006debye,groma2007dynamics,groma2010variational} 
because of the stress screening observed by 
DDD simulations for close to neutral systems ($\kappa=\rho_+-\rho_-$ is much 
smaller than $\rho=\rho_++\rho_-$)
correlations can be well accounted for by adding a quadratic term in  $\kappa$ 
to $P_{\rm sc}$. With this term, the potential reads as
\begin{eqnarray}
P[\chi,\rho_{\pm}]= P_{\rm sc}[\chi,\rho_{\pm}]+ P_{\rm 
corr}^{\pm}[\chi,\rho_{\pm}]
\end{eqnarray}
where
\begin{eqnarray}
  P_{\rm corr}^{\pm}[\chi,\rho_{\pm}]=\int \frac{T_0}{2}\frac{\kappa^2}{\rho} 
{\rm d}x{\rm d}y,
  \label{eq_P} 
\end{eqnarray}
in which  $T_0$ is a constant (with the dimension of force) determined by the dislocation-dislocation 
correlation function \cite{groma2007dynamics}. With the phase field formalism for conserved quantities 
the evolution equations for the fields $\rho_{\pm}$ take the form
\begin{equation}
 \dot{\rho}_{\pm} +\partial_x j_{\pm}=0 \ \ {\rm with} \ \ j_{\pm}= \mp 
M\rho_{\pm} \partial_x \frac{\delta P}{\delta \kappa}.
 \label{eq_evol}
\end{equation} 
It should be mentioned that since the dislocation system is not a 
thermodynamical one there is no a priori reason that a phase field approach can 
be applied. So, the correctness of the above form has to be justified. Comparing 
it with the field equations obtained earlier \cite{groma2003spatial} by a systematic coarse-graining 
procedure of the ``discrete'' system of evolution equations (\ref{eq_motion}), 
one can see, that the phase field Eq. (\ref{eq_evol}) is indeed justified if 
$|\kappa|/\rho \ll 1$ \cite{groma2007dynamics}. 

For many configurations, like close to a grain boundary, however, the  
$|\kappa|/\rho \ll 1$ condition, a key assumption in the microscopic derivation 
of the continuum theory, is not fulfilled. Therefore a new concept is needed to 
construct the correlation term.  The primary aim of the present paper is to 
formulate a phase field theory if only one type of dislocation is present (say 
$\rho_+$), representing the other extreme case $|\kappa|/\rho=1$ . (As it is 
discussed later, the general $|\kappa|/\rho$ case can be established from the 
two extremes in a straightforward manner.)  

Since $P_{\rm sc}[\chi,\rho_+]$ represents the mean field (i.e. correlationless) 
term,  it is not affected by the  
$|\kappa|/\rho$ ratio. The real nontrivial question is the possible form of 
$P_{\rm corr}[\chi,\rho_+]$ in this case. As a first possible approximation one 
can look for a term that does not contain the spatial derivatives of $\rho_+$. 
From simple dimensionality considerations the general form of such a term is
\begin{equation}
 P_{\rm corr}[\rho_+]=\int T \rho_+ {\rm f}(\rho_+/\rho_0){\rm d}x{\rm d}y,
\end{equation}
where T is a constant, ${\rm f}(x)$ is an arbitrary function and $\rho_0$ is a parameter with 
inverse length square dimension.  For the following consideration a key point to 
notice is that since the dislocation-dislocation interaction is scale free, i.e. 
it does not contain any length scale parameter, the evolution equation of 
$\rho_+$ also cannot contain any parameter with length dimension but the local dislocation 
spacing. As a consequence of 
this, the form of ${\rm f}(x)$ has to be chosen so that $\rho_0$ does not appear 
in the phase field equation (\ref{eq_evol}). To fulfill this condition the only 
possibility is if ${\rm f}(x)\propto\ln(x)$. With the above 
form of $f(x)$ Eq. (\ref{eq_evol}) takes the form
\begin{equation}
 \dot{\rho}_++Mb\partial_x\left\{ \rho_+\left [ \tau_{\rm sc} - \frac{T}{b\rho_+} 
\partial_x \rho_+ \right ]\right\}=0, \ \ 
  \tau_{\rm sc}=\partial_x  \partial_y\chi
 \label{eq_d}
\end{equation}
where $\tau_{\rm sc}$ is the ``self consistent'' or ``mean 
field'' shear stress. The evolution equation 
(\ref{eq_d}) has to be supplemented with appropriate boundary conditions. This 
depends on the actual properties of the boundaries, but it is quite a common 
case that the boundary is unpenetrable for the dislocations, so the dislocation 
current has to vanish at the boundaries if the Burgers vector is not parallel to 
the surface.

One can easily see, however, that the above ``diffusive'' like evolution 
equation is not satisfactory. Namely, let us consider a channel with surfaces 
perpendicular to the dislocation glide direction. After randomly placing 
dislocations with the same Burgers vectors into the channel and allowing the 
system to relax, a DDD simulation shows that the system does not remain 
homogeneous, boundary layers develop at the surfaces. A 
typical relaxed dislocation configuration obtained by DDD can be seen in Fig. 1, 
while the dislocation density obtained by averaging 5000
different realizations is plotted in Fig. 2.  

\begin{figure}[ht]
 \includegraphics[width=4cm,angle=-90]{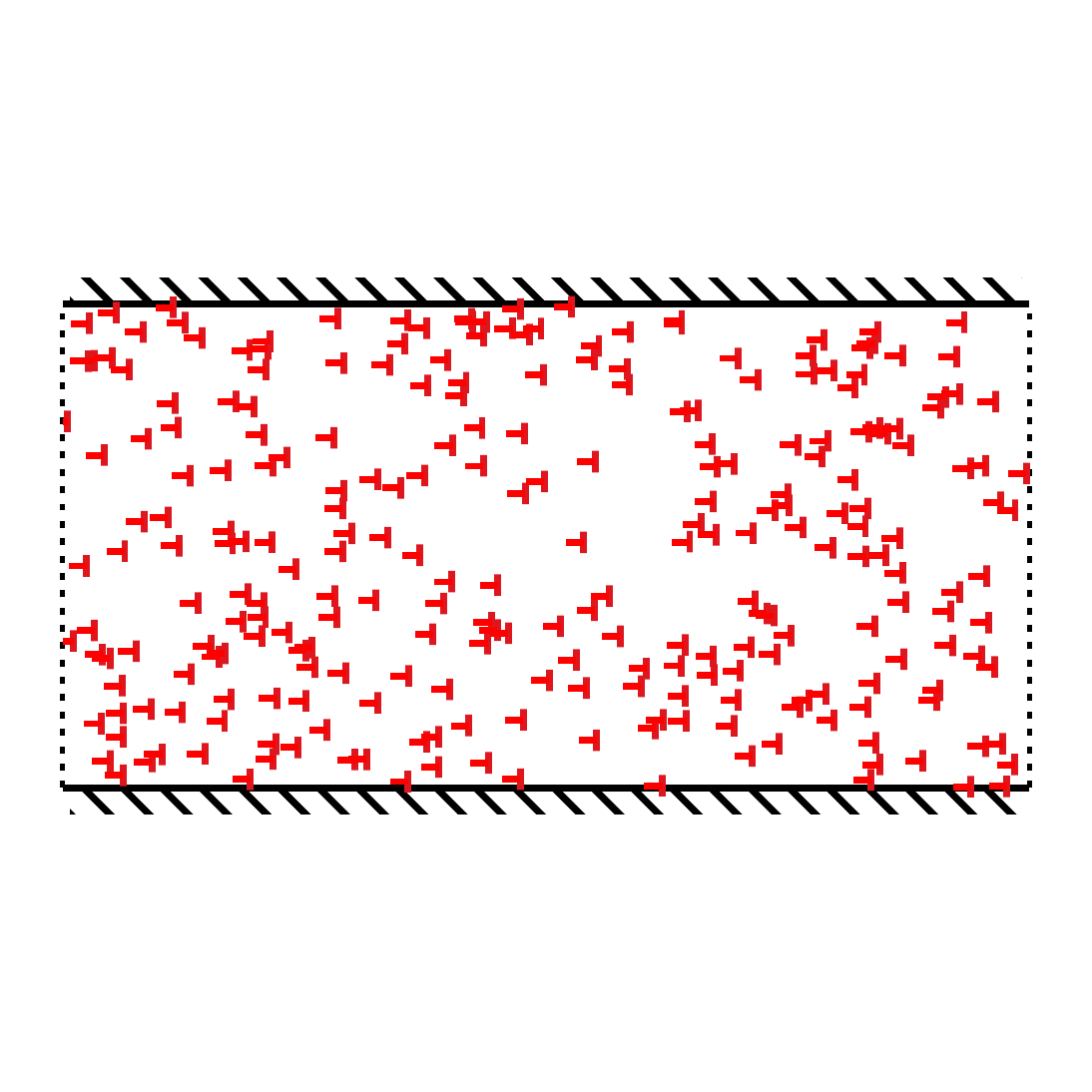} 
\hspace{5pt}
 \includegraphics[width=4cm,angle=-90]{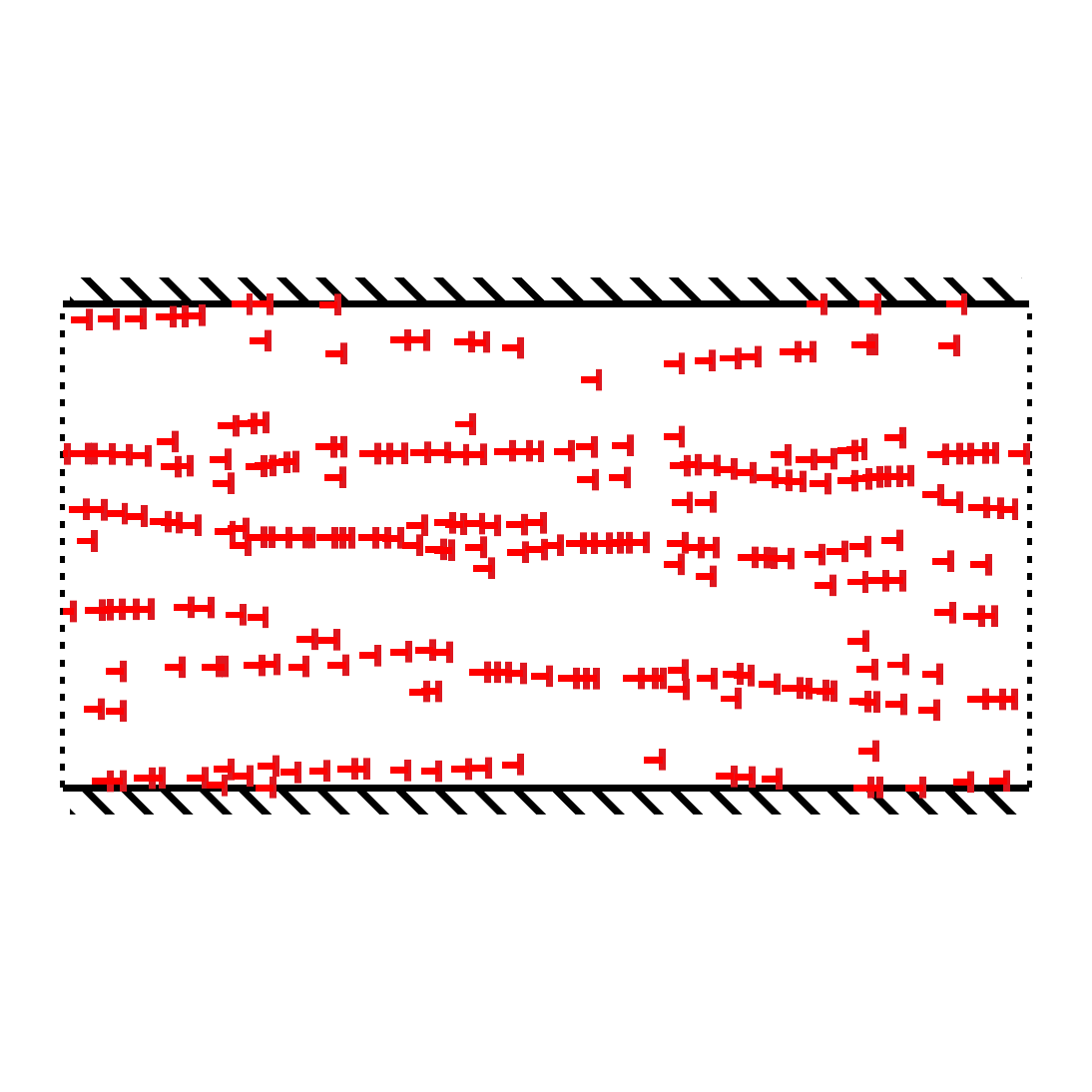}
 \caption{Random initial configuration of dislocations in a channel (left box). 
Relaxed dislocation configuration (right box). The walls are unpenatrable and  
periodic boundary condition is used in the $y$ direction. The total number of 
dislocations is 256.}  
\end{figure}

\begin{figure}[ht]
 \begin{center} \includegraphics[width=5cm,height=4cm,angle=0]{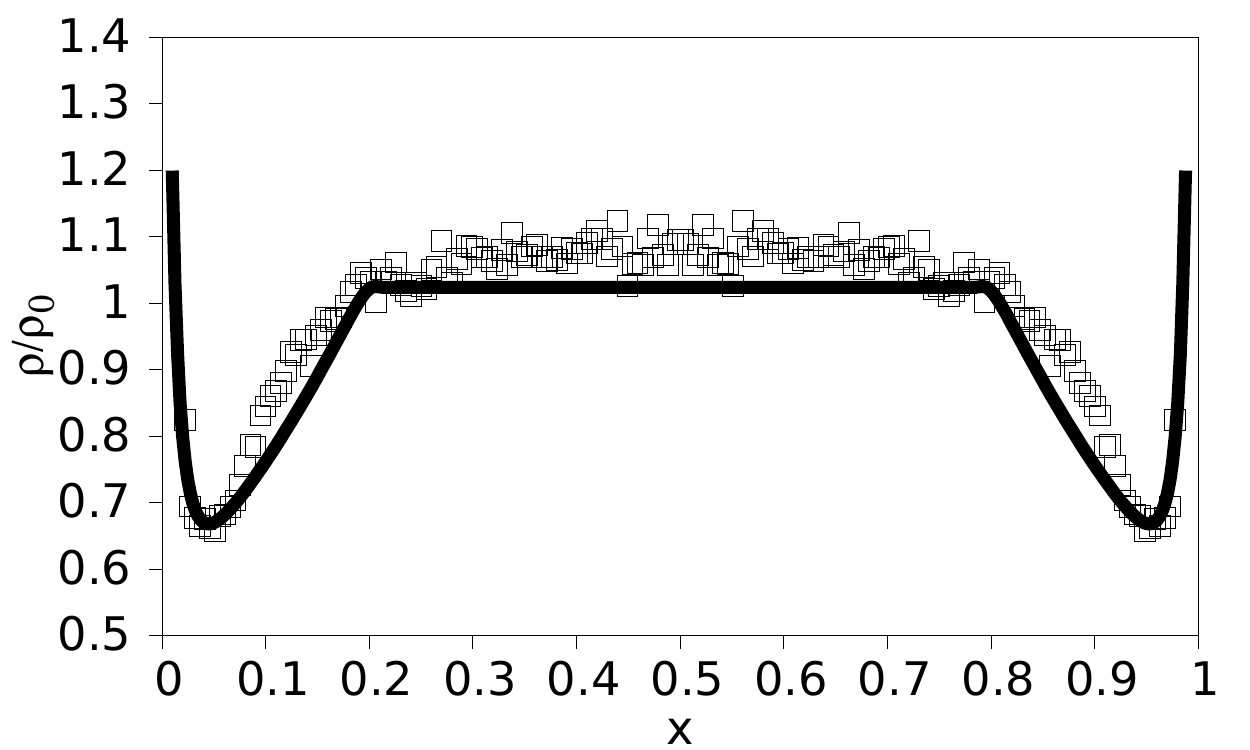} 
\end{center}
 \caption{Dislocation density profile (relative to the initial density), 
 averaged in the direction perpendicular  to the slip direction, developing between two 
 unpenetrable walls after the 
relaxation of an initially random system of dislocations with the same Burgers 
vectors obtained by DDD simulation (squares). The numerical solution of the phase 
field model proposed (full line). Relevant simulation parameters are: $u$=0.1, 
$W_x$=5.5, $\alpha_m=0.1$. }
 \label{fig_pde}
\end{figure}

On the other hand, however, in case of zero external shear stress the 
homogeneous $\rho_+$ is a stable solution of Eq. (\ref{eq_d}) obtained above.  
So one can conclude, Eq. (\ref{eq_d})  is not able to reproduce the dislocation 
configuration developing in a channel.  As the form of $ P_{\rm corr}$ is 
dictated by the scale free nature of dislocation-dislocation interaction, to 
resolve the discrepancy between the DDD simulation results and the prediction of 
Eq. (\ref{eq_d}) one has to introduce gradient terms in $\rho_+$ into $ P_{\rm 
corr}$. Again, to avoid the appearance of length scale parameters in the 
evolution equation the possible form 
of $ P_{\rm corr}$ depending on $\bigtriangledown \rho_+$ is 

\begin{equation}
 P_{\rm corr}=\int_D T \rho_+ \left [\ln\left (\frac{\rho_+}{\rho_0}\right ) + 
{\rm u}\left(\frac{\bigtriangledown \rho_+ \hat{S}\bigtriangledown \rho_+ 
}{2\rho_+^3} \right) \right ]{\rm d}x{\rm d}y,
 \label{eq_Pcorr}
\end{equation}
where $\hat{S}$ is a symmetric dimensionless 2x2 matrix and ${\rm  u}(x)$ is an 
arbitrary function. If $|\bigtriangledown \rho_+/\rho_+^{3/2}|\ll 1$ one can 
take the leading linear term in ${\rm  u}(x)$, so $ P_{\rm corr}$
used in the considerations below is quadratic in $\bigtriangledown \rho_+$. 

Due to the gradient terms introduced in  $ P_{\rm corr}$ the phase field 
equation (\ref{eq_evol}) is a fourth order
partial differential equation in $\vec{r}$. In order to get unique solution 
further boundary conditions have to be introduced beside the one introduced 
earlier for the dislocation current $\vec{j}_+$. A dislocation wall developing 
next to a boundary has an extra surface energy which can be accounted for by 
adding a surface term to $ P_{\rm corr}$. For dimensionality reasons the surface 
energy density has to be proportional to $\sqrt{\rho_+}$, but as above, 
parameters with length scale should not be introduced in the evolution equation 
of the dislocations, so the only possible form of the surface $(\partial D)$ 
contribution to the potential $P$ is
\begin{equation}
 P_{\rm sf}[\rho_+]=\oint_{\partial D} \alpha_{\rm sf} T \sqrt{\rho_+}\vec{n} {\rm d} 
\vec{A}
 \label{eq_Psf}
\end{equation}
where the $\vec{n}=\vec{b}/b$ term takes into account that in the surface energy 
only  the surface projection perpendicular to the slip plane has contribution, 
and $\alpha_{\rm sf}$ is a constant. (One may consider an appropriate 
$\bigtriangledown \rho_+$ dependence of $\alpha_{\rm sf}$ but in this paper only 
the leading term independent from $\bigtriangledown \rho_+$ is taken.) Since the 
relaxation of the dislocation configuration next to the surface is expected to 
be much faster than in the bulk, the boundary condition can be obtained from the 
total plastic potential 
\begin{equation}
P[\chi, \rho_+]=P_{\rm sc}[\chi, \rho_+]+P_{\rm corr}[\rho_+]+ P_{\rm 
sf}[\rho_+]
\end{equation}
given by Eqs. (\ref{eq_Pm},\ref{eq_Pcorr},\ref{eq_Psf}) as
\begin{equation}
 \left . \frac{\delta P}{\delta \rho_+} \right |_{\partial D}= \left 
.\vec{W}\bigtriangledown \rho_+-\rho_+^{3/2} \right | _{\partial D}=0 
\label{eq_bc2}
\end{equation}
where $\vec{W}$ is a dimensionless constant 2D vector depending on $\hat{S}$, 
$\alpha_{\rm sf}$ and the surface direction.  

The system of Eqs. (\ref{eq_evol},\ref{eq_bc2}) together with the condition that 
$j_+$ vanishes at the system surface, represent a closed set of equations with 
unique solution. As it is discussed below, however, it is not able to account 
for the dislocation density evolution obtained by DDD simulation for the channel 
problem mentioned above.  Namely, for this geometry due to the translation 
symmetry in the $y$ direction Eq. (\ref{eq_evol}) has a steady state solution 
satisfying the condition
\begin{equation}
 \frac{\delta P}{\delta \rho_+}=\mu_0, \label{eq_st}
\end{equation}
where $\mu_0$ is a parameter (analogous to the chemical potential) depending on 
the initial average dislocation density. After substituting the actual form of 
$P[\chi, \rho_+]$ given by Eqs. (\ref{eq_Pm},\ref{eq_Pcorr},\ref{eq_Psf}) into 
Eq. (\ref{eq_st}) we arrive at a second order ordinary differential equation for 
the steady state $\rho_+(x)$. With the analysis of the structure of the equation 
one can find that within the channel the steady state solution is either 
completely convex or concave depending on the actual value of the parameters, 
i.e. it is not able to recover the shape seen in Fig. 2. even for a general 
$u(x)$. Another related issue is that according to DDD simulation results, the 
relaxed configuration of a dislocation system can vary if the initial 
dislocation density field is rearranged while the total number of dislocations 
(or average dislocation density) is kept constant.  So, the steady state 
dislocation  density does not reach always the same configuration represented by 
Eq. (\ref{eq_st}) at the same physical parameters. 

To resolve the problem it is natural to assume that a system of dislocations 
with identical sign has an ``internal rigidity'' meaning that if the internal 
shear stress $\tau_{\rm int}=\partial_x (\delta P /\delta \rho_+) -\tau_{\rm 
sc}$ is smaller than a critical value the system cannot rearrange itself. This 
is somewhat similar to the ``flow stress'' of neutral systems but for a single 
signed system the flow stress is obviously zero since under an external stress 
the whole system can move rigidly. Although the ``internal rigidity'' is a 
dislocation-dislocation correlation effect (like the flow stress introduced in 
\cite{groma2003spatial} for neutral systems) there is no trivial way to take it into account by 
adding an appropriate term to $P[\chi, \rho_+]$. Within the phase field 
framework, however, it is possible to introduce a mobility function giving the 
dislocation current as
\begin{equation}
 j_+=Mb\rho_+ \left[M(\tau_{\rm int})+ \tau_{\rm sc} \right] \label{eq_jm}
\end{equation}
with
\begin{equation}
 M(\tau)=\left\{
 \begin{array}{ll}
   0 \ \ & {\rm if} \ \ \tau<\tau_0 \\
   \tau-\tau_0 \ \ &{\rm if} \ \ \tau>\tau_0
 \end{array}
  \right .  
 \label{eq_M}
\end{equation}
Since there is no other length scale but the dislocation spacing, from a simple 
dimensionality consideration $\tau_0=\alpha_{\rm m}bD^{-1}\sqrt{\rho_+}$. The 
quantity $\alpha_{\rm m}$ may depend on the possible different dimensionless 
combinations of the  dislocation density and its derivatives but in our analysis 
it was kept constant. An important consequence of this ``critical type'' 
mobility function is that the dislocation system cannot reach the configuration 
corresponding to the minimum condition given by Eq. (\ref{eq_st}).   

As it is seen in Fig. \ref{fig_pde}  for the channel problem the numerical 
solution of the evolution equation with Eqs. 
(\ref{eq_evol},\ref{eq_jm},\ref{eq_M}) recovers the characteristic feature of 
the spatial variation of the dislocation density obtained by DDD.

Besides the channel problem discussed above it is interesting to analyze what 
happens with a localized dislocation density ``peak'' formed from dislocations 
with the same Burgers vector and homogeneous in the $y$ direction.  According to DDD 
simulation results,  if one considers a dislocation density peak with random 
dislocation positions it starts to spread out but it reaches a steady state 
shape depending on the initial width and dislocation density. Without the  
gradient term in $P_{\rm corr}$ given by Eq. (\ref{eq_Pcorr}) and the nontrivial 
mobility function (\ref{eq_M}) the evolution equation would be a ``diffusion'' 
like equation predicting a complete spread out of the density peak. So, the main 
message of this paper is the non-diffusive behavior of the dislocation system.  

According to the discussion explained above we have at hand continuum theories 
of dislocations in two extreme cases: if $|\kappa/\rho|\ll 1$ and if 
$|\kappa/\rho|=1$. It is natural to assume that the general $\kappa/\rho$ case 
can be obtained by a smooth interpolation between the limits. (Since the mean 
field part of the plastic potential $P_{\rm sc}$ is valid for any $\kappa$ we 
have to consider only the correlation part of $P$).  As a first step let us 
simply take the sum  $P^t_{\rm corr}[\rho_+,\rho_-]= P_{\rm corr}[\rho_+]+P_{\rm 
corr}[\rho_-]$. If $|\kappa/\rho| \ll 1$ and we neglect the terms depending on 
the derivatives of the dislocation densities one obtains that $P_{\rm 
sc}+P^t_{\rm corr}$ recovers the form of $P$ given by Eq. (\ref{eq_P}) if 
$T=T_0$. Since, however, $T$ and $T_0$ are determined by the 
dislocation-dislocation correlation functions \cite{groma2006debye,groma2007dynamics}
depending on the $\kappa/\rho$ 
ratio,  one cannot expect that $T=T_0$. It is usefull to rewrite, however, the two 
logarithmic terms in $P^t_{\rm corr}$ into 
the form
\begin{eqnarray}
 && T\rho_+\ln(\rho_+/\rho_0)+T\rho_-\ln(\rho_-/\rho_0) \nonumber \\  
&&=\frac{T}{2}\rho\ln\left[\frac{\rho^2-\kappa^2}{4\rho_0^2}\right]+\frac{T'}{2}
\kappa\ln\left[\frac{\rho+\kappa}{\rho-\kappa}\right] \label{eq_pp}
\end{eqnarray}
(with $T=T'$). For a general $\kappa/\rho$ the coefficient $T'$ can have a weak 
$\kappa^2/\rho^2$ dependence in the form of $T'(x)=T+(T_0-T)(x-1)^2/2$. (Since 
in the evolution equations the functional derivative has to be taken only with 
respect to $\kappa$, term depending only on $\rho$ can be dropped out from 
$P^t_{\rm corr}$.) Two things that should be mentioned at this point: 
{\it i)} the $\frac{dT'}{dx}(1)=0$ condition ensures that no extra terms appears 
in $\delta P/\delta \kappa$ at $|\kappa|=\rho$ discussed above. {\it ii)} the 
coefficient in front of the first term in the right hand side of Eq. 
(\ref{eq_pp}) has to remain $\kappa^2/\rho^2$ independent to ensure the $\rho_0$ 
does not appear in the evolution equation of the dislocation densities. Without 
going into the details we mention a weak $ \kappa^2/\rho^2$ dependence of the 
coefficient in front of the gradient term in Eq. (\ref{eq_Pcorr}) can be 
introduced in a similar way. Certainly the actual values of the parameters 
appearing in the general form of $P_{\rm corr}$ have to be determined from DDD 
simulations corresponding to different system geometries. 

In summary, a continuum theory of straight parallel dislocations is proposed 
that takes into account dislocation-dislocation correlation effects. The theory 
is obtained from a functional of the dislocation densities by applying the 
formalism of phase field theories. Although the phase field functional is 
established on a phenomenological ground, the actual form of the functional is 
largely dictated by the scale free nature of the dislocation-dislocation 
interaction. The theory is validated by comparing its predictions with DDD simulation.
It has to be stressed that the form of the phase field functional proposed is the simplest 
possible one (containing only 
the leading order terms)  that is able to recover the characteristic feature of 
the DDD simulation results. In order to recover the fine details of the DDD 
simulation results one may have to introduce higher order terms. Furthermore, 
certainly the 2D dislocation geometry the continuum theory is corresponding to 
is a strong 
simplification of the real much more complex 3D ones. In the 3D continuum   
theory, however,  the structure of the terms corresponding to the correlation 
between dislocation loops should have rather similar forms.

\begin{acknowledgments}
 Financial supports of the Hungarian Scientific Research Fund (OTKA) under 
contract numbers K-105335  and PD-105256
and of the European Commission under grant agreement No. CIG-321842 are also 
acknowledged. 
\end{acknowledgments}

\end{document}